\documentstyle[11pt,newpasp,twoside,epsfig]{article}
\newcommand{\gcc}{\ \rm{g\ cm^{-3} }}
\newcommand{\micros}{\ \mu s}
\newcommand{\cms}{\ \rm{cm \ s^{-1}}}

\newcommand{\avgsub}[2]{\left<#1\right>_{#2}}

\def\edcomment#1{\iffalse\marginpar{\raggedright\sl#1\/}\else\relax\fi}
\reversemarginpar

\begin{document}
\title{Investigations of Pointwise Ignition of Helium Deflagrations on Neutron Stars}

\author{M.~Zingale, S.~E.~Woosley, A.~Cumming}
\affil{University of California Santa Cruz, Santa Cruz, CA, 95064}
\author{A.~Calder, L.~J.~Dursi, B.~Fryxell, K.~Olson, P.~Ricker,
        R.~Rosner, F.~X.~Timmes}
\affil{University of Chicago, Chicago, IL 60637}

\begin{abstract}
We look at the spreading of localized temperature perturbations in the
accreted fuel layer of a non-rotating neutron star.  The pressure at
the base of the accreted fuel layer is large and the material is only
partially degenerate.  Any temperature perturbations and resulting
pressure gradients will lead to enormous accelerations (both laterally
and vertically) of the material in the fuel layer.  If the burning of
this fuel cannot proceed more rapidly than the spreading of this
perturbation, then localized burning cannot take place, and it is
likely that the ignition would have to proceed simultaneously
throughout the envelope.  We present some multidimensional simulations
of the spreading of temperature perturbations in a helium atmosphere
on a neutron star.
\end{abstract}

\section{Introduction}

We wish to investigate whether localized burning can occur in the
accreted atmosphere of a neutron star.  A summary of the observations
of type I X-ray bursts is presented by Cumming (2002) in this
proceedings.  Fryxell \& Woosley (1982) made some estimates of how fast
a deflagration can travel through a fuel layer, considering
multidimensional effects.  We look to see whether it is possible to
setup such a deflagration to begin with.  Given a temperature
perturbation at the base of the fuel layer, a pressure gradient will
result, which will act to smooth out this initial perturbation.  This
gradient will drive velocities which will spread out the initial
perturbation.  We want to determine whether it is possible for nuclear
burning to act on a timescale that is faster than this spreading
timescale.

\section{Nuclear Timescales}

\begin{figure}[t]
\begin{center}
\epsfig{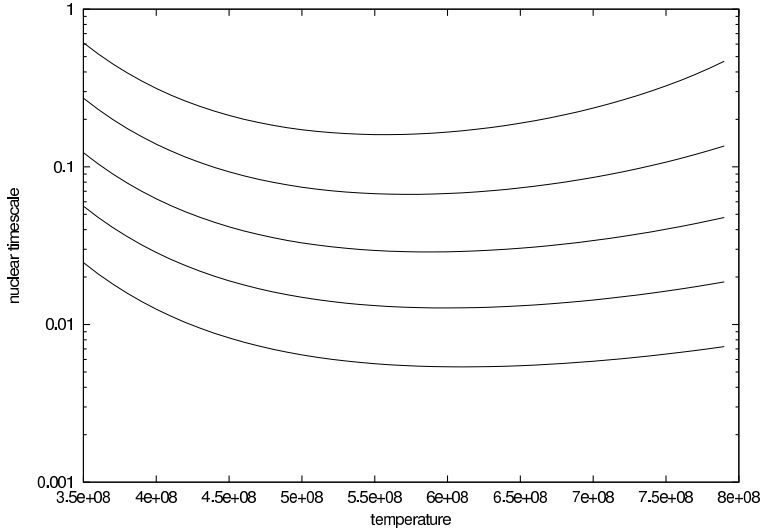}
\vskip -0.25in
\end{center}
\caption{\label{fig:nuc} Nuclear timescales vs
temperature for 5 different densities (from top to bottom, $5\times
10^5 \gcc$, $7.1\times 10^5 \gcc$, $1\times 10^6\gcc$, $1.4\times
10^6\gcc$, and $2\times 10^6\gcc$) for the 3-$\alpha$ reaction.}

\end{figure}

The nuclear rise time for a region of fuel is computed as
\begin{equation}
\tau_{\rm{nuc}} = \int_{T_{\rm{fuel}}}^{2T_{\rm{fuel}}}
      \frac{c_p \rm{d}T}{e_{\rm{nuc}} {\dot X}_{\rm{fuel}}}
\end{equation}
If we cannot burn faster than the spreading time of a perturbation,
then the burning will not be localized.  Figure \ref{fig:nuc} shows
the nuclear timescales for several different densities of pure helium
as a function of temperature, using the 3-$\alpha$ reaction.  We see
that the fastest burning timescale is $\sim 0.08$s----this is long
compared to the dynamical timescales that we consider in the next
section.

We want how long it takes for two regions of fuel (hot and cool) to 
show differences---we want to compare the rates of the burning
\begin{equation}
\Delta \tau_{\rm{nuc}} = \frac{1}{1/\tau_{\rm{hot}} - 1/\tau_{\rm{cool}}}
\end{equation}
Two regions of fuel that have nearly the same temperature will have very close
nuclear rise times ($\tau_{\rm{hot}} \sim \tau_{\rm{cool}}$), so
$\Delta \tau_{\rm{nuc}}$ will be large---it takes a long time to see 
any differences. 

We can ignite locally if the time for the temperature
perturbation to spread away is longer than $\Delta
\tau_{\rm{nuc}}$.  For each density, we need to find the initial
conditions (if they exist) that result in a spreading time that is
slower than this nuclear timescale.

\section{\label{spread_timescale} Estimates of the Spreading Timescale}

Any temperature gradient on the grid will produce a pressure gradient,
that will lead to a lateral acceleration.  The acceleration laterally 
through the atmosphere can be integrated to yield an estimate for the spreading velocity,
\begin{equation}
a_x = \frac{{\bf \nabla}P \cdot {\bf \hat x}}{\rho} \Longrightarrow
        v_x \approx c_s \left ( {\frac{\delta P}{P}} \right )^{1/2}
\end{equation}
An initial 20\% temperature perturbation is a $\sim 6\%$ pressure
perturbation (at a density of $2\times 10^6 \gcc$), which leads to a
velocity of $\sim 7\times 10^7 \cms$.  If our initial perturbation is
1000 cm across, this would spread in $\sim 15\micros$.  Note that
$\delta P$ decreases as the material spreads, so this is the fastest
case.

Rotation can have a stabilizing effect on the spreading of an initial
perturbation (Spitkovsky, Levin, \& Ushomirsky 2002).  There is a
scale (Rossby deformation length) at which the Coriolis force will balance a pressure gradient:
\begin{equation}
L = \sqrt{g H_0} / f
\end{equation}
where f is the Coriolis force and $H_0$ is the scale
height. Perturbations smaller than this scale will spread until they
reach this length.  For a 300 Hz rotation speed, this scale is about
10 km.  This means that any temperature perturbation will spread out
until it reaches 10 km.  This allows us to define what we mean by
non-rotation---the maximum size for $L$ is the circumference of the
neutron star, so if the rotational frequency is less than $\sim 100 \
\rm{Hz}$, then rotation will not matter.  This is the regime that
we are considering presently.

\section{Initial Conditions}

\begin{figure}[t]

\begin{center}
\epsfig{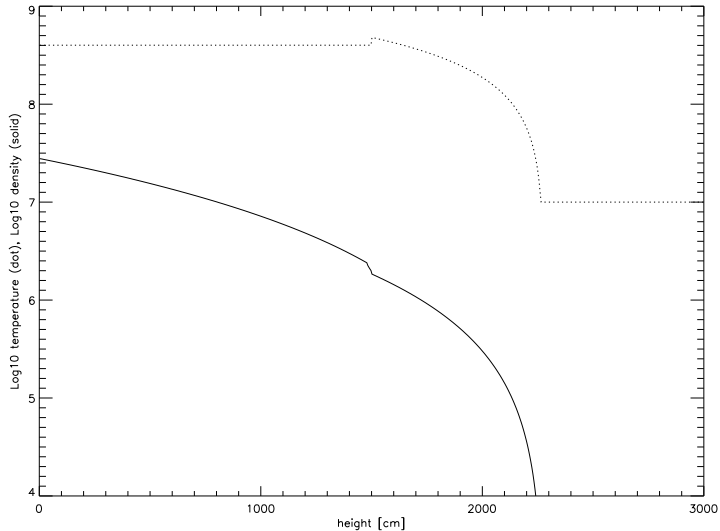}
\vskip -0.25 in
\end{center}

\caption{ \label{1d_models} Initial model in the maximally perturbed
region, showing temperature and density.  This model is in hydrostatic
equilibrium (HSE) vertically.  The temperature perturbation is placed at the
fuel/ash interface ($1500 \ \rm{cm}$).  The density, temperature,
and composition in the zone immediately below the interface are
identical in all models, so the models have equal pressures at all
heights lower than the interface.}

\end{figure}

The initial model we use to study the spreading of a perturbation
through a neutron star atmosphere consists of a fuel and ash region,
in HSE vertically, and in lateral pressure
equilibrium in the ash region.  It is impossible to put an initial
model in HSE vertically and pressure equilibrium
laterally if the temperatures at the bases differ.

The initial conditions are created by computing separate
initial models for each column of zones on the grid.  The fuel 
is pure $^4$He with a (possible) temperature perturbation at its base.
The ash is pure $^{56}$Ni and completely isothermal.  The model is
completely specified by the ash temperature ($T_{\rm{ash}}$), the
temperature and density at the base of the fuel layer, and the height
of the interface.

For the present simulations, we choose the density at the base of the
fuel layer to be $2\times 10^6 \gcc$.  The temperature in the fuel
region is specified by
\begin{equation}
\frac{d T}{d z}  = C \cdot \frac{\delta_P g}{c_P}; \quad
\delta_P = - \left ( \frac{\partial \ln \rho}{\partial \ln T} \right )
\end{equation}
and $C < 1$ is a constant multiplier (we choose 0.9, making it just
sub-adiabatic). We pick an ash temperature of of $T_{\rm{ash}} = 4
\times 10^8 \ \rm{K}$.
\begin{figure}[t]

\begin{center}
\epsfig{file=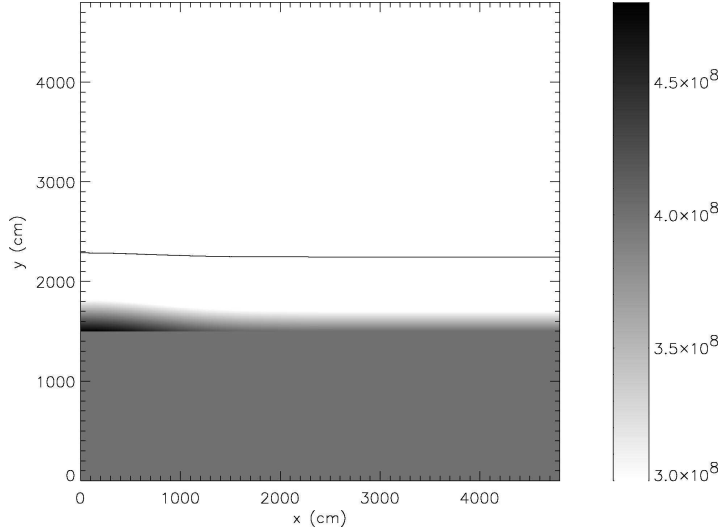, height=2.75in}
\vskip -0.25 in
\end{center}
\caption{\label{temp_init} Initial temperature perturbation.  The
black contour marks a density of $10\gcc$, indicating the top of
the initial atmosphere.}

\end{figure}
The temperature at the base of the fuel layer is perturbed with a Gaussian
profile,
\begin{equation}
\label{eq:temp_perturb}
T(y) = T_{\rm{ash}} + (T_{\rm{perturb}} - T_{\rm{ash}}) \cdot
	\exp\{-(y/L)^2\}
\end{equation}

We integrate the equation of HSE outward from the
fuel/ash interface using a second-order polynomial reconstruction for
the zone-averaged density and a third-order reconstruction for the
pressure:
\begin{equation}
\label{eq:second_order_diff}
\avgsub{P}{+1}-\avgsub{P}{0}  = 
\frac{g \delta}{12} \left ( 5\avgsub{\rho}{+1} + 8\avgsub{\rho}{0}
                  - \avgsub{\rho}{-1} \right )
\end{equation}
where $\delta$ is the zone width, and $\avgsub{P}{0}$ represents the
average value of the pressure in zone 0.  This reconstruction greatly
suppresses any transients when an unperturbed model is evolved on our
grid.

Figure \ref{1d_models} shows the vertical structure of the initial
models at the maximally perturbed base temperature.  Each column of
our computational domain is given a unique initial model, according to
the fuel base temperature specified by Eq.\ (\ref{eq:temp_perturb}).
The material up to the base of the fuel layer is in lateral pressure
equilibrium.  Figure \ref{temp_init} shows the temperature
perturbation in the initial model (with a peak temperature of
$4.8\times 10^8 \ \rm{K}$) We note that there is only a pressure
gradient in the $y$ direction in the fuel layer.

\section{Computational Method}

The calculations were performed with the FLASH code (Fryxell
et al. 2000), solving the Euler equations with the piecewise-parabolic
method (PPM) (Colella \& Woodward 1984).  Cylindrical geometry
($r$, $z$) were used to exploit the axisymmetric symmetry of the
problem.  Modifications to PPM were made to better evolve a
hydrostatic atmosphere, and are discussed in Zingale et
al. (2002).  We use a Helmholtz free energy based tabular equation of
state (Timmes \& Swesty 2000), which allows for degenerate/relativistic electrons.  No burning was used for these
calculations---we only wish to determine the hydrodynamical effects.

The accurate solution of the spreading problem requires us to extend
the physics of the atmosphere into the boundary conditions.  In the
absence of any perturbations, the Euler equations reduce to the
condition of hydrostatic equilibrium.
When the flow is hydrostatic, the boundary conditions are very
influential.  We use a hydrostatic boundary at the base of our model
that provides pressure support to the material above it, while
allowing sound waves to flow out.  The material velocities are
reflected in the boundary and the temperature is held constant.  The
density and pressure in the boundary are computed using the
second-order zone-averaged differencing of HSE Eq.\
(\ref{eq:second_order_diff}) coupled with the EOS (see Zingale et
al. 2002).

At the upper boundary, we extend the domain for several 10s of meters
above the surface of the atmosphere, to allow for the differences in
the pressure scale heights of the perturbed and unperturbed regions.
The density of this material above the atmosphere is small so as not
to be dynamically important.

\section{Results}

\begin{figure}[t]

\begin{center}
\epsfig{file=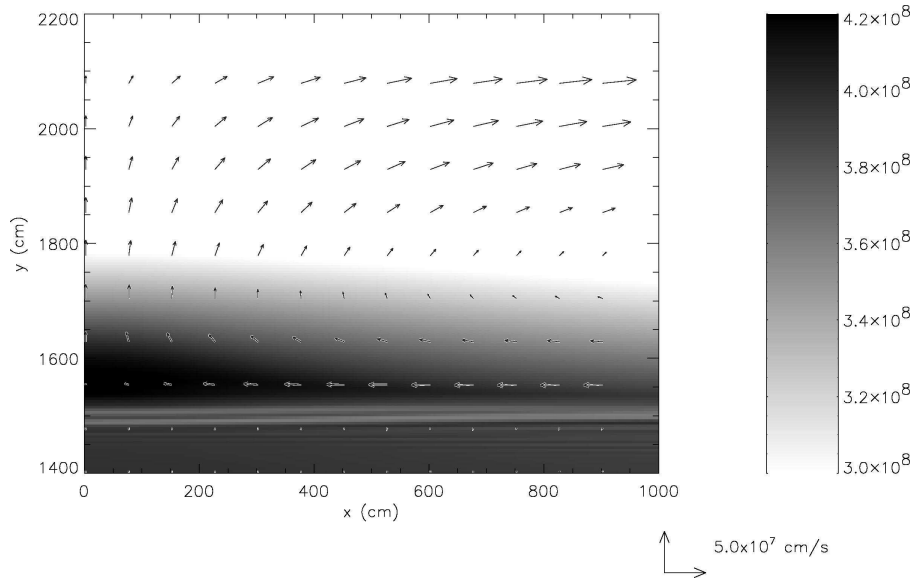, height=3.0in}
\vskip -0.25 in
\end{center}
\caption{\label{fig:vel} Temperature (grayscale) and velocity (vectors)
structure after 30 ns for the 20\% perturbation.}

\end{figure}

\begin{figure}[t]

\begin{center}
\epsfig{file=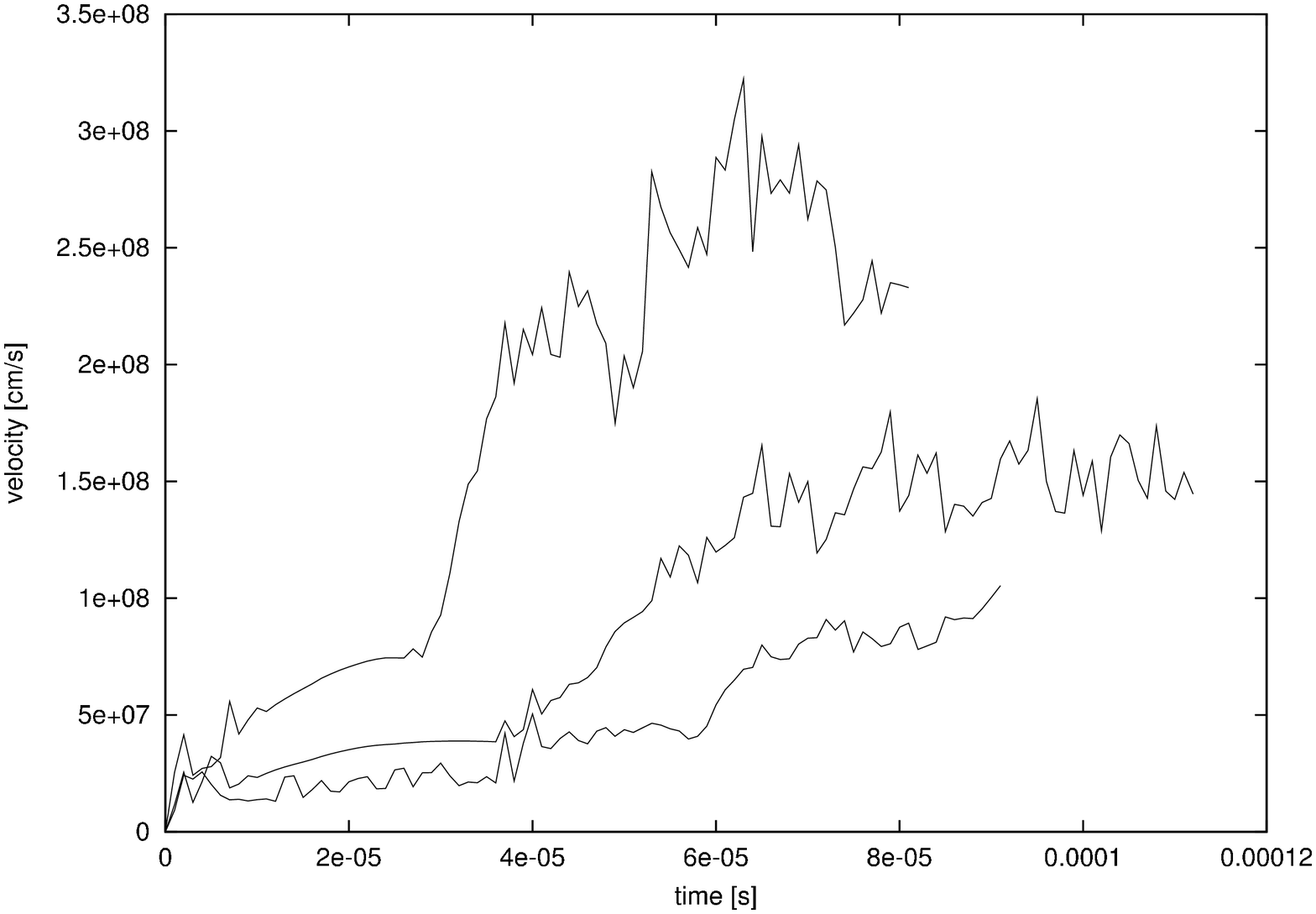, height=2.75in}
\vskip -0.25 in
\end{center}
\caption{\label{vel_time} Maximum lateral velocity in the atmosphere
as a function of time for the 3 different perturbations, 5\% (lower),
10\% (middle), and 20\% (upper).}

\end{figure}

We ran 3 different temperature perturbation simulations, 5\%, 10\%,
and 20\% (corresponding to 1.6\%, 3.2\%, and 6.4\% pressure
perturbations).  Shortly after the simulation begins, a roll develops,
with material comings towards the symmetry axis on the bottom, and
flowing away on the top.  This is shown in Figure \ref{fig:vel} for
the 20\% temperature perturbation.  The maximum spreading velocity we
find as a function of time is shown in Figure \ref{vel_time}.  We see
that, as expected, the spreading speed is greatest for the largest
perturbation.  It is difficult to determine from the data whether it
follows the $(\delta P)^{1/2}$ dependence that we estimated in \S
\ref{spread_timescale}

\begin{figure}[h]

\begin{center}
\epsfig{file=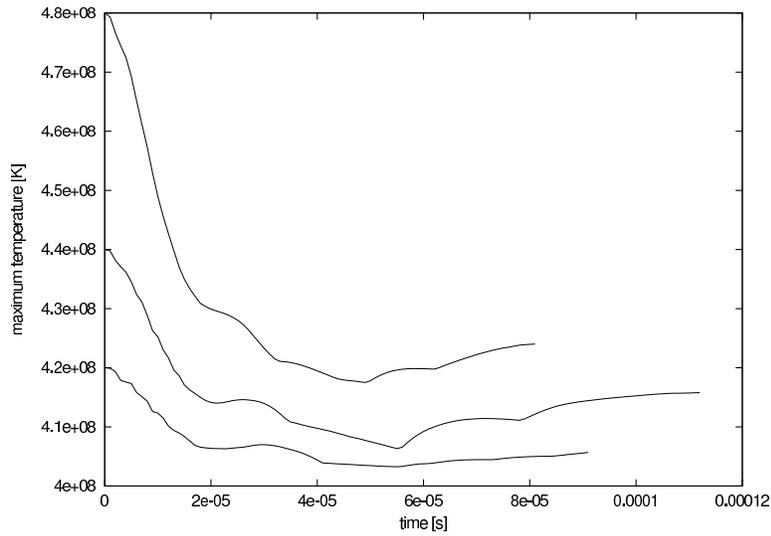, height=2.75in}
\vskip -0.25 in
\end{center}
\caption{\label{temp_time} Maximum temperature on the grid as a
function of time for the 3 different initial models, 5\% (lower), 10\%
(middle), and 20\% (upper).}

\end{figure}

Figure \ref{temp_time} shows the maximum temperature on the grid as a
function of time for the 3 runs.  We expect the burning to be most
vigorous at the peak temperature on the grid at any time, so this is
in effect a measure of the nuclear timescales.  Recall that in all
cases, the ambient temperature is $4\times 10^8 \ \rm{K}$.  We see
that in all cases, the temperature decreases very rapidly down to a
level slightly higher than the background.  In some instances, it even
appears to increase slightly at long times---the reason for this is
not known presently.  The timescales of the decrease is $\sim 10s
\micros$---much faster than the nuclear timescale, and in agreement
with our estimates in \S \ref{spread_timescale}

All of these estimates are much faster than the burning timescale, so
this perturbation is unlikely to result in localized ignition.  We can
try longer perturb\-ations---but the maximum length is the
circumference of the star.  Work is underway to generalize these
results for different densities and perturbation sizes to determine
whether a set of initial conditions exist under which we can locally
burning the accreted fuel on a non-rotating neutron star.

\ \\ These simulations were carried out with FLASH 2.1
(http://flash.uchicago.edu/).  Support for this work was provided by
the Scientific Discovery through Advanced Computing (SciDAC) program
of the DOE, grant number DE-FC02-01ER41176.  Some calculations we
performed on the NERSC IBM SP/3 (seaborg) at LBNL.

\begin{references}

\reference Colella, P. \& Woodward, P.~R. 1984, JCP, 54, 174
\reference Cumming, A. 2002, this proceedings.
\reference Fryxell, B. et al. 2000, \apjs, 131, 273
\reference Fryxell, B. A. \& Woosley, S. E. 1982, 261, 332
\reference Spitkovsky, A., Levin, Y., \& Ushomirsky, G. 2002, \apj, 566, 1088
\reference Timmes, F.~X. \&  Swesty, F.~D. 2000, \apjs, 126, 501
\reference Zingale, M. et al. 2002, \apjs, accepted

\end{references}
\end{document}